# Study of $^3$He inelastic scattering on $^{13}$C and $^{14}$C at 37.9 MeV

M. N. El-Hammamy[1], A. Attia[2], F.A. El-Akkad[2] and A. M. Abdel-Moneim[2]

[1] *Physics Department, Faculty of Science, Damanhur, Damanhur University, Egypt*
[2] *Physics Department, Faculty of Science, Alexandria, Alexandria University, Egypt*

**Abstract** The differential cross sections of elastic and inelastic scattering of $^3$He ions on $^{13}$C and $^{14}$C have been studied at energy of 37.9 MeV with a double folding model based on M3Y-Reid effective nucleon-nucleon interaction. The resulted parameters have been used for the standard Distorted Wave Born Approximation calculations of angular distributions corresponding to different excitations levels of $^{13}$C and $^{14}$C and deformation parameters have been deduced.

## 1. Introduction

Folding model optical potential to describe the elastic and inelastic scattering of nuclear particles has become widely accepted. In the folding model the optical and transition potentials are calculated by folding the ground state density distribution of the projectile $\rho(r)$ and the transition density of the target nucleus $\rho_{tr}(r)$, respectively, with a suitable nucleon-nucleon (NN) interaction. The transition density is obtained either from the nuclear structure calculation or by deforming the ground state density distributions [1-3]. Such calculations are used to analyze nuclear reactions and are compared with the experimental data and phenomenological calculations [4-7]. Successful description of angular distributions of elastic scattering cross section for light heavy colliding nuclei at several energies were obtained using double folding (DF) alpha cluster potentials [8,9]. Various values of renormalization constant for the real part of the optical potential to obtain a good fits to the experimental data were used.

The studies of inelastic scattering have most often used a transition form factors (FF) to calculate the inelastic scattering observables, e.g., in Distorted Wave Born Approximation (DWBA).

Several versions of the effective NN interaction (e.g. JLM [10], M3Y [11]) have been tested for fitting data of elastic and inelastic scattering systems. In this work Reid-M3Y effective NN interaction with Zero Range Approximation is used. The DF- FF are calculated with two forms, the volume and surface WS. The resulting parameters are used in the elastic and inelastic scattering analysis simultaneously.

The aim of the present work is to derive the FF for the reaction $^3$He – $^{13}$C and $^{14}$C on physical basis by folding the $^3$He density into $^{13}$C and $^{14}$C density and an integral form of NN interaction. The derived FF is used to analyze the elastic and inelastic scattering of $^3$He with an energy of 37.9 MeV leading to the excitation of $^{13}$C levels, 3.68 MeV (3/2$^-$), 3.85 MeV (5/2$^+$), 8.84 MeV (1/2$^-$), 11.85 MeV (7/2$^+$) [12] and $^{14}$C levels 6.73

MeV($3^-$), 7.01 MeV($2^+$), 8.32 ($2^+$) MeV [13]. We analyzed inelastic scattering to known low-lying states to test the ability of calculations to predict the angular distributions and the strength of these states. Our analysis includes the study of the differences between parameters of the two scattering systems, the change of deformation parameters with number of neutrons.

The method employed here and discussions are given in Section 2, and the conclusions are presented in Section 3.

## 2. Analysis and discussion

To study the elastic scattering for the reactions of $^3$He - particles with $^{13}$C and $^{14}$C, the program code *DFPOT* [14] has been used. $V_F$ (r) is the DF potential carried out by introducing the effective NN interaction over the ground state density distribution of the two colliding nuclei. It is evaluated from the expression

$$V_F(r) = \iint \rho_1(r_1) \rho_2(r_2) V_{NN}(s) dr_1 dr_2 .  \quad (1)$$

$\rho_1(r_1)$ and $\rho_2(r_2)$ are the density distribution of the two colliding nuclei, and $V_{NN}(s)$ is the effective NN interaction potential, which is taken to be a standard Reid- M3Y interaction [4] in the form,

$$V_{NN}(r_{12}) = 7999.0 \frac{e^{-4.0r}}{4.0r} - 2134.0 \frac{e^{-2.5r}}{2.5r} + J_{00}(E)\delta(r). \quad (2)$$

The first term represents the direct part and the second term represents the exchange part of the interaction potential. It plays an important role in reproducing the experimental results for elastic and inelastic scattering [8,9]. The exchange part can be written to a good approximation in the form [4]

$$J_{00}(E) = -276(1 - 0.005 \frac{E}{A}) , \quad (3)$$

where E is the energy in the center of mass system and A is the mass number of the projectile. In our calculations, the nuclear matter density distribution of 3He nucleus has the form $\rho = \rho_\circ \exp(-\alpha r^2)$ where $\alpha = 0.5505 \; fm^{-2}, \rho_\circ = 0.2201 \; fm^{-3}$ [15], and it is assumed the modified oscillator density parameters for both $^{13}$C and $^{14}$C compiled in reference [16].

The total optical potential must comprise both the real part and the imaginary part, the latter being responsible for the absorption of the incident particle in the inelastic channels. Since the M3Y interaction is real, the folding calculation gives the real part of the optical potential. In the model used here, the volume imaginary part is taken with the WS shape and a surface absorptive term proportional to its derivative as follow ,

$$W(r) = \frac{-W_V}{(1+\exp(\frac{r-R_V}{a_V}))} + \frac{4W_S \exp(\frac{r-R_S}{a_S})}{(1+\exp(\frac{r-R_S}{a_S}))^2} \quad , \quad R_{V,S} = r_{V,S}(A_T^{1/3} + A_P^{1/3}) \quad (4)$$

Thus, the total potential is

$$U(r) = \left[ N_r V_F(r) - \varphi_r \frac{dV_F(r)}{dr} \right] + iW , \quad (5)$$

where, $N_r, \varphi_r, W_V, W_S, r_V, r_S, a_V, a_S$ are variable parameters and $V_F$ (r) is the double folding potential of equation (1), The surface term is added into the real part of the potential, imitating the contribution of the dynamic polarization potential [17,18].

The folded FF obtained in each case is introduced into the program code *DWUCK4* [19] to compute the differential scattering cross section.

The calculations for elastic scattering was calculated by DWUCK4. *Firstly*, we used [(WS) + (SWS)] for real parts of the potential as case one. Comparisons are shown in figure (1) between the present calculations and experimental data.

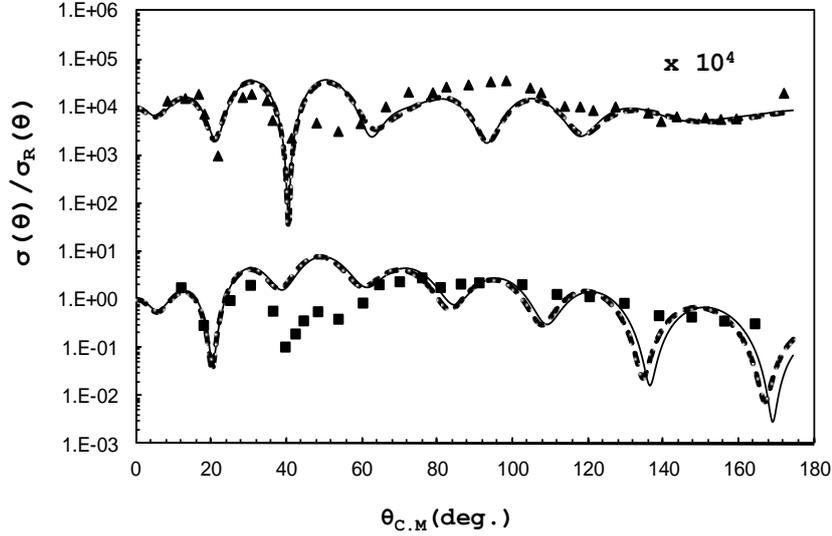

**Figure(1).** Angular distributions of $^3$He elastically scattered on $^{14}$C and $^{13}$C. The theoretical cross section obtained with DF model is represented by *dashed line* for case one and *solid lines* for case two. Experimental points are denoted by *black symbols*, ▲ for $^{13}$C and ■ for $^{14}$C.

It is well to see that, for a light nuclei-target, the theoretical to experimental data are not satisfactory adjusted. Thus, in a region of angles up to 60 degrees the theoretical sections, well describe the experimental data. This discrepancy of theoretical and experimental data at larger angles, apparently means that other reaction mechanisms begin to play an essential role. This is in accordance with the results from reference [20], in which nuclei-targets with A = 12-50 and at energy 50.5 MeV, values of the $\varphi_r$ parameter are equal to zero i.e. it doesn't have effect, but for larger masses it has either in elastic or inelastic cases. In our work $\varphi_r = 0.091$ for $^{13}$C and 0.087 for $^{14}$C in case one.

Therefore *secondly in case two*, we neglect this parameter (real SWS is equal to zero), as we expect it will not have an effective role. The comparison between the two cases is shown in figure (1). The variable parameters are listed in table (1).

***Table (1):*** *Elastic real folded normalization parameters ($N_r$ and $\varphi_r$), real volume integrals ($J_V$) and imaginary WS parameters for volume and surface parts.*

| Reaction | $N_r$ | $J_V$ MeV.fm$^3$ | $W_V$ MeV | $r_V$ fm | $a_V$ fm | $4W_S$ MeV | $r_S$ fm | $a_S$ fm | $\varphi_r$ |
|---|---|---|---|---|---|---|---|---|---|
| $^3$He + $^{13}$C | 0.58 | 234.45 | 2.39 | 1.33 | 0.572 | 25.71 | 1.19 | 0.876 | 0 |
| $^3$He + $^{14}$C | 0.59 | 238.49 | 1.88 | 1.66 | 0.614 | 7.34 | 1.216 | 0.878 | 0 |

The difficulty found in fitting elastic scattering cross sections is reflected in the inelastic predictions and indicates a deficiency in the present form of the potential.

The analysis of the inelastic scattering of the $^3$He particles has been performed and the comparison of theoretical calculations and the experimental data has been presented for final states up to 12 MeV excitation in figure (2).

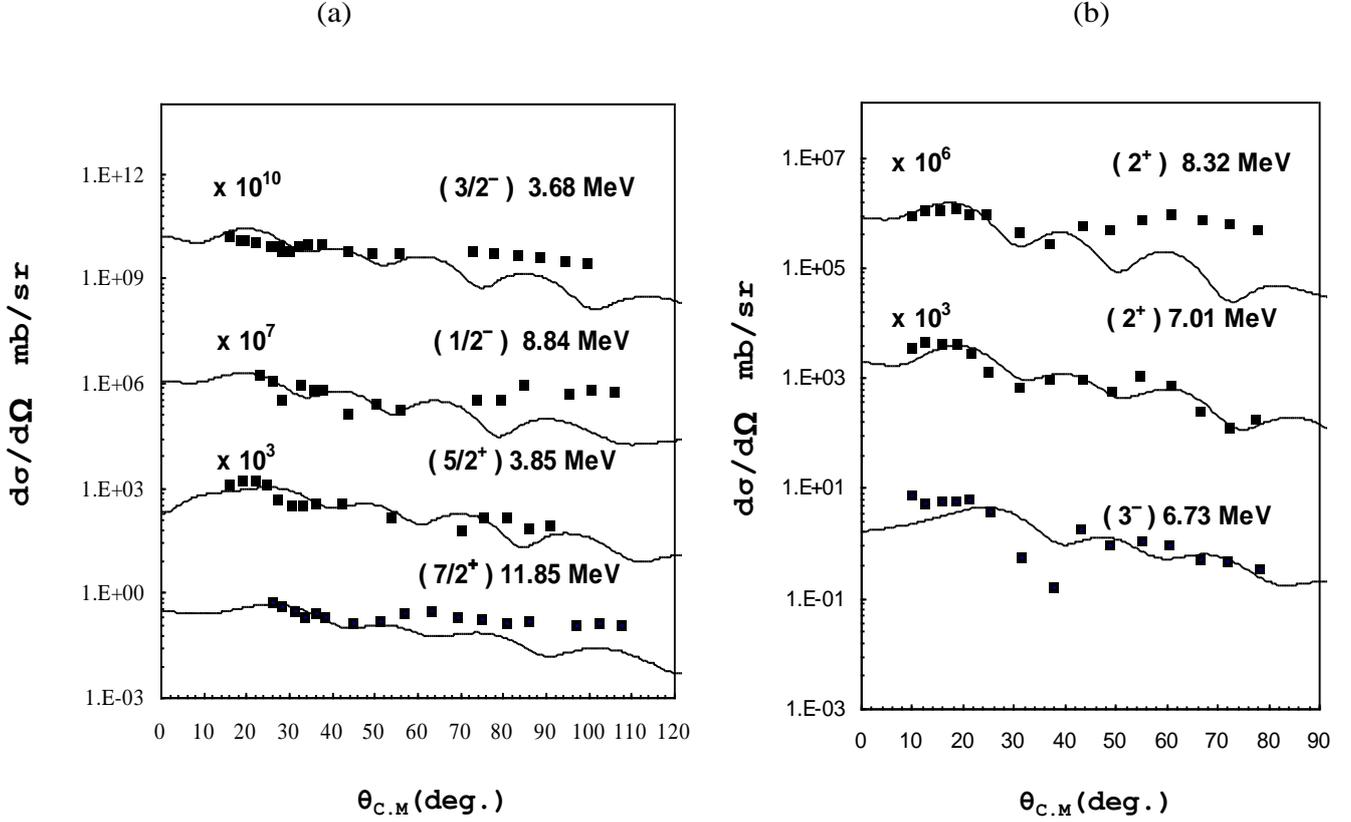

**Figure (2).** Angular distributions of $^3$He inelastically scattered on $^{13}$C (a) and $^{14}$C (b) compared with DF model predictions (solid lines).

The fitted FF normalization parameters with real volume integrals and imaginary WS parameters [4] listed in table (1) are used to calculate the inelastic scattering potential by calculating the derivative of the real elastic potential multiplied by a deformation parameter $\beta_\lambda$, where λ is the multi-polarity. So the transition potential for a given excited state is

$$V^\lambda(r) = -\beta_\lambda \frac{d}{dr} V(r). \qquad (6)$$

Here, the transition form factors for the inelastic scattering of $^3$He to the different states in $^{13}$C and $^{14}$C are calculated using the folding model together with real deformation parameters ($\beta_r$) which produce the best fit of the inelastic scattering data for real equivalent folded transition FF respectively.

In order to estimate the quality of the fit, one can calculate a relative error

$$\chi_R^2 = \frac{1}{N} \sum_{i=1}^{i=N} \left[ \frac{(\sigma^{calc.}(\theta_i) - \sigma^{exp.}(\theta_i))}{(\sigma^{calc.}(\theta_i) + \sigma^{exp}(\theta_i))} \right]^2, \qquad (7)$$

and an alternative option is to minimize an absolute error

$$\chi_A^2 = \frac{1}{N}\sum_{i=1}^{i=N}\left[(\sigma^{calc.}(\theta_i) - \sigma^{exp.}(\theta_i)\right]^2 . \qquad (8)$$

Where N is the number of differential cross section data points and $\sigma^{calc.}(\theta_i)$ is the $i^{th}$ calculated cross section and $\sigma^{exp.}(\theta_i)$ is the corresponding experimental cross section. The extracted deformation parameters and as well as the corresponding values of $\chi_R^2$ are shown in table (2).

**Table (2):** Deformation parameters and $\chi_R^2$ values from best fit to the *inelastic scattering data for different levels of $^{13}C$ and $^{14}C$.*

| Reaction | Level | $\beta$ | | | $\chi_R^2$ |
|---|---|---|---|---|---|
| $^3$He + $^{13}$C | 3.58 (5/2$^+$)<br>11.85(7/2$^+$) | 0.09<br>0.07 | | | 0.11<br>0.20 |
| | 3.68 (3/2$^-$)<br>8.84 (1/2$^-$) | 0.11<br>0.03 | | | 0.21<br>0.35 |
| $^3$He + $^{14}$C | 6.73(3$^-$)<br>7.01(2$^+$)<br>8.32(2$^+$) | 0.46<br>0.09<br>0.03 | | 0.20-0.27 [13]<br>0.19-0.25 [13]<br>0.09-0.15 [13] | 0.1<br>0.04<br>0.24 |

In case of $^{14}$C, the deformation parameters extracted from the present calculation are different with those obtained by the analysis of the inelastic scattering of $^3$He from $^{14}$C in reference [13].The equivalent folded transition FF leads in most of cases to smaller cross sections. There are some differences between experimental and calculated values even at forward angles at the level of 6.73(3$^-$) MeV, according to reference [13],as it is above the threshold for a neutron emission. But it gives better agreement with experimental data at levels 7.01 (2$^+$), 8.32 (2$^+$) MeV and a worse adjustment at the level of 6.73(3$^-$) MeV than reference [13].

In case of $^{13}$C, the shapes of the resulting angular distributions did not change significantly than that of reference [12]. However, the model describes worse the angular distributions which is clearly found at level 3.68 MeV (3/2$^-$).

## 3 .Conclusion

The use of real folded FF in calculations is proved to be adequate to describe the existing experimental data for $^3$He-particles elastic and inelastic scattering with a minimal number of fitting parameters.

The final potential gives good description for the elastic scattering cross-section in forward area of angles. For large angles the correspondence was not reached, what can be probably explained by the contribution to elastic scattering of exchange mechanisms resulted in back scattering. Also, it gives a satisfactory description of the inelastic cross section.

In conclusion, one can notice that the description of the overall experimental data is at least not worse than achieved in others work, but here in all cases we used the same M3Y effective interaction, and the nucleon densities for all nuclei were treated in a unified frame work. Consequently, similar microscopic studies could be immensely helpful for nuclear structure and reaction studies with the availability of a larger data set on different nuclei.

We expect that the fit with experimental data can be improved if we use the exact exchange part of the effective NN interaction in calculating the real part instead of the zero range which we have used in the present calculations. Also, the results improvement can be achieved if we use another fitted potential form as $WS^2$ or using different forms of density dependence which increase the quality of fit.